          [2001/04/25 1.1 (PWD)]
\documentclass[a4paper,twocolumn]{esapub} 
\usepackage{natbib}
\usepackage{graphicx}
\usepackage{epsfig}

\title{The INTEGRAL 
View of the Galactic Nucleus}

\author[1,2]{A. Goldwurm}
\author[1,2]{G. B\'elanger}
\author[1,2]{P. Goldoni}
\author[1,2]{J. Paul}
\author[1,2]{R. Terrier}
\author{M. Falanga}
\affil[1]{Service d'Astrophysique, DAPNIA/DSM/CEA - Saclay,
91191 Gif-sur-Yvette, France}
\affil[2]{F\'ed\'eration de Recherche Astroparticule et Cosmologie, 
11 Pl. Berthelot, 75005 Paris, France}

\author[3]{P. Ubertini}
\author[3]{A. Bazzano}
\author[3]{M. Del Santo}
\affil[3]{IASF-CNR, via del Fosso del Cavaliere 100, 00133 Roma, Italy}

\author[4]{C. Winkler}
\author[4]{A.N. Parmar}
\author[4]{E. Kuulkers}
\affil[4]{Research and Scientific Support Department, ESA ESTEC,
PB 299, NL-2200 AG Noordwijk, Netherlands}

\author[5]{K. Ebisawa}
\affil[5]{Integral Science Data Center, Chemin d'Ecogia, 16,
CH-1290 Versoix, Switzerland}

\author[6]{J.P. Roques}
\author[6]{G. Skinner}
\affil[6]{Centre d'Etude Spatiale des Rayonnements, CNRS,
9, avenue du Colonel Roche, Toulouse Cedex 4, France}

\author[7]{N. Lund}
\affil[7]{Danish Space Reasearch Institute, DK-2100 Copenhagen 0,
Denmark}

\author[8]{F. Melia}
\affil[8]{Physics Dept. and Stewart Obs., University of Arizona,
1118 E. 4$^{\rm th}$ St., POB 210081, Tucson, AZ 85721, US}

\author[9]{F. Yusef-Zadeh}
\affil[9]{Dept. of Physics and Astronomy, Northwestern University, Evanston, IL 60208, US}

\def\etal{et~al.~}

\def\sgra{Sgr~A$^*$}

\def\int{INTEGRAL}

\begin{document}
\maketitle

\begin{abstract}
We present the preliminary results of the
observational campaign performed in 2003 to study the
Galactic Nucleus with INTEGRAL.
The mosaicked images obtained
with the IBIS/ISGRI coded aperture instrument in
the energy range above 20 keV,
give a yet unseen view of the high-energy sources
of this region in hard X- and gamma-rays, with
an angular resolution of 12$'$.
We report on the discovery of a source,
IGR J17456-2901, compatible with the
instrument's point spread function and
coincident with the Galactic Nucleus \sgra to within 0.9$'$.
The source is visible up to 60-80~keV with a
20--100~keV luminosity at 8~kpc of
$3 \times 10^{35}$~erg~s$^{-1}$.
Although we cannot unequivocally associate
the new INTEGRAL source to the Galactic Nucleus,
this is the first report of significant
hard X--ray emission from within
the inner 10$'$ of the Galaxy and
a contribution from the galactic center
supermassive black hole itself cannot be excluded.
Here we discuss the results obtained
and the perspectives for future observations of
the Galactic Nucleus with INTEGRAL and other observatories.
\end{abstract}
\section{Introduction}
The Galactic Nucleus (GN) is among the most interesting objects in the
high-energy sky, as it links our own galaxy with active galactic nuclei
and quasars.
At a distance of 8~kpc (Eisenauher et al. 2003), it hosts the closest massive
black hole (BH) to us, as shown in particular by near-infrared (NIR)  observations
of star proper motions near the galactic center (GC) \cite{sch02,gen03,ghe03}.
The detailed  NIR observations of the central cluster of young and luminous stars
have now constrained the enclosed mass within 0.001 pc ($\sim$~120~AU)
to a value which implies a core density $>$~10$^{17}$~M$_{\odot}$~pc$^{-3}$.
This density is most naturally explained by the presence of
a massive black hole of $(4.0 \pm 0.8)$~10$^{6}$~M$_{\odot}$.
Such a BH would have a ~Schwarzschild
radius ($R_\mathrm{S}$) of about 1.2~10$^{12}$cm or 0.08 AU and is supposed
to accrete the environmental matter producing detectable emission in
a broad frequency range \cite{mel01}.
\\
The bright ($\sim1$~Jy), compact, non-thermal radio source Sgr~A$^*$,
discovered exactly 30 years ago by Balick and Brown (1974)
and located at less than 0.01$''$ (83 AU) from the dynamical center of the
central star cluster is believed to be the radio conterpart of this extreme object.
Its radio spectrum is described by an inverted or flat power law with  high and low
frequency cut-offs, and a peculiar sub-mm bump at frequencies $>$~100~GHz.
Only recently detected in NIR both in quiescent,
and flaring states \cite{gen03,ghe04}, this source is undetectable
in the visible and UV bands due to the large absorption and is also very weak in X-rays.
The Chandra Observatory measured in 1999 a quiescent luminosity of only
$L_\mathrm{X}\mathrm{[2-10\ keV]}\approx2\times10^{33}$~erg~s$^{-1}$ from
\sgra\ and the emission appeared partly extended.
However, in October 2000, Chandra detected a bright 3~hr flare, characterized by
rapid intrinsic variability \cite{bag01,bag03}. During this flare
the luminosity increased to
$L_\mathrm{X} \mathrm{[2-10\ keV]}\approx10^{35}$~erg~s$^{-1}$ in 4~ks,
the power law spectrum hardened, with a change of the photon index from
2.7 in quiescence to 1.3, and a rapid decrease on a timescale
of 600~s was observed, thus implying an emitting region of
size $<$~20~$R_\mathrm{S}$.
Two other bright X-ray flares from Sgr~A$^*$ were detected with XMM-Newton
on September 4, 2001 and on October 3, 2002 \cite{gol03a,por03}.
In 2001 XMM-Newton detected the beginning of a flare in the last 900 s
of an observation pointed towards the GN
(Goldwurm et al. 2003a). In this interval the \sgra\ flux
increased by up to a factor 30 and the spectrum hardened to a slope of index 1.
In October 2002 the most powerful flare from this source was discovered
with XMM-Newton. It lasted only 2.7 ks but the source luminosity reached
$\approx$~3.6~$\times$~10$^{35}$~erg~s$^{-1}$
with an increase factor of nearly 200 in luminosity and a rather soft spectrum
($\alpha  \approx 2.5$) \cite{por03}.
\\
These last observations have opened challenging new questions regarding the
accretion process, activity and emission mechanisms at work
in Sgr A* and have also re-opened
the possibility of observing the source in the hard X-ray and gamma-ray bands ($>$ 10-30 keV).
At energies between 10 and 20 keV a number of detections of Sgr~A$^*$ were claimed in the past,
the most significant were those based on Spacelab 2 observations in 1985 \cite{ski87}
and those of  ART-P on GRANAT in 1990-1991 \cite{pav94}.
Skinner et al. (1987) reported the detection with the XRT/SL2 of a source compatible with
\sgra\ and noted that it was much brighter than the extrapolation of the soft X-ray emission
seen with the Einstein Observatory. The ART-P data also showed
presence of emission in the 8-20 keV band at even higher flux levels, impling that
the source 4-20 keV luminosity ranged between 5 and 10 $\times 10^{35}$~erg~s$^{-1}$
(see Goldwurm 2001 for a review).
The identification with ASCA (2-30 keV) of a transient eclipsing binary at $\approx 1.3'$
from \sgra\  in 1994,  the source AX~J17456-2901 (Maeda et al. 1996, Sakano et al. 2002),
casts some doubt on the association of the high energy detections with the GN,
because the associated error boxes ($\sim$ 3$'$) included this transient source.
The coded mask gamma-ray telescope SIGMA on the GRANAT satellite
performed a deep $9 \times 10^{6}$~s survey of the
central parts of the galaxy between 1990 and 1997,
but could provide only upper limits for
the hard X- and $\gamma$-ray emission from the neighbourhood of \sgra\
at energies above 35~keV \cite{gol94,goi99}.
\\
The derived low bolometric luminosity of the GN,
in contrast with the powerful output from active galactic nuclei
or black hole binaries,
has motivated the development of several models for
radiatively inefficient accretion onto or ejection from the
central supermassive black hole \cite{nar98,fal00}.
These models have been then widely applied
to other accreting systems but their validity for \sgra\ have been challenged
by more and more precise measurements over the last
10 years \cite{gol01,mel01,bag03}.
The observed X-ray flares and the
very recent discovery with the VLT NACO imager \cite{gen03}
and the Keck telescope \cite{ghe04},
that \sgra is also the source of frequent IR flares
could indicate the presence of
an important population of non-thermal electrons in the
vicinity of the black hole \cite{mar01,liu02,yua02,yua03,liu04}.
These results have raised great interest in the possibility
of observing hard X-rays from the GN,
a measure of which may particularly shed light
on the relative role of accretion and ejection
in the \sgra\ system.
\\
We have recently analyzed the large set of data
collected with INTEGRAL during the galactic center survey performed in the
first part of 2003, and
found some excess emission at energies $>~$20 keV from the region including the Sgr~A
complex \cite{bel04}.
We review these first INTEGRAL results on the GN and present a more recent
analysis performed on the entire 2003 INTEGRAL data set.
\begin{table*}[htbp]
\begin{center}
\caption{List of the 2003 INTEGRAL observations of the Galactic Nucleus (GN)}\vspace{1em}
\begin{tabular}[h]{lccccc}
\hline \hline
\\
Observation &  Target  & Mode                      &      Dates   (2003)     &  Exposure     &  GN Eff. Exp. \\
 Type$^a$   &          &                           &  Start~~~~~~~End          &  (ks)       &        (ks)      \\
\\
ToO & XTE~J1720--318    & 5~$\times$~5 D      & 28/02~ - ~02/03    & 176           & 128 \\
GCDE & Survey           & Survey           & 02/03~ - ~01/05     & 675           &369 \\
ToO &  H~1743--31       & 5~$\times$~5 D   & 06/04~ - ~22/04     & 280          & 256 \\
GO  &  \sgra            & 5~$\times$~5 D   & 30/08~ - ~24/09   &  1000        & 938 \\
GCDE & Survey           & Survey          & 02/08~ - ~14/10    &  675          & 374 \\
\hline \hline
\end{tabular}
\end{center}
Notes: \\
a) GO = Guest Observer Observation, GCDE = Galactic Center Deep Exposure (Core Program),
ToO= Target of Opportunity Observation \\
b) 5~$\times$~5 D = 5~$\times$~5 Dithering Pattern, 7 HEX = 7 Hexagonal Dithering Pattern
\label{tab:log}
\end{table*}

\section{Observations and Data Analysis}

INTEGRAL (INTErnational Gamma-Ray Astrophysics Laboratory),
the ESA gamma-ray observatory \cite{win03} launched on 2002 October 17,
carries two main instruments, the gamma-ray imager IBIS and
and the gamma-ray spectrometer SPI,
and two monitors, JEM-X, the Joint European X-ray Monitor,
and OMC, the Optical Monitoring Camera.
The results reported here were
obtained with the IBIS coded mask imaging instrument (Ubertini et al. 2003),
sensitive over the energy range between 15~keV and 10~MeV and
characterised by a wide field of view (FOV) of
$29^{\circ} \times 29^{\circ}$,
an angular resolution (FWHM) of 12$'$ and a sensitivity of about 1~mCrab
at 100~keV for 1~Ms exposure.
The IBIS performance in the low energy range (15--1000 keV)
is achieved thanks to the ISGRI camera (Lebrun et al. 2003)
made up of more than 16000 CdTE detectors.
\\
The Galactic Center is a priority target for the INTEGRAL mission.
A specific survey of the central regions of the galaxy, the
Galactic Center Deep Exposure (GCDE) program, is performed each year with
the goal of mapping these regions
($-30^{\circ} < l < 30^{\circ}$, $-15^{\circ}< b < 15^{\circ}$) at high energies \cite{win01}.
A first set of GCDE observations was performed between March and May 2003
and another set between August and October.
The GN was in the IBIS FOV also during two Target of Opportunity (ToO) observations
dedicated to the X-ray novae H~1743-31 (Parmar et al. 2003) and XTE~J1720-318
(Cadolle Bel et al. 2004).
Moreover a dedicated observation of 1~Ms,
to search for a gamma-ray counterpart to \sgra\ was performed in September 2003.
Table~1 summarizes the characteristics of these observations which provided the data
used in this work. The total effective exposure on \sgra, accounting for
partial coding and performed data selections, was $\sim$~2.1~Ms.
\\
INTEGRAL observations are generally made of several exposures,
performed at fixed pointing directions with a specific pattern on the
plane of the sky and each having durations ranging from 1800~s to 4000~s
(Courvoisier \etal 2003).
The reduction and analysis of the IBIS/ISGRI data were performed with
the INTEGRAL Offline Scientific Analysis (OSA) package provided by the
INTEGRAL Science Data Center (ISDC).
The algorithms relative to the IBIS data analysis are described in
Goldwurm et al. (2003b).
\\
In a preliminary study we treated only the set of data collected between 2003 February 28
and May 1, using the OSA 2.0 version of the analysis software and without
correction for background structures. To account for the residual systematic noise
we measured the distribution of the residuals in the images and applied a correction
factor to reduce the significance accordingly.
The results published by B\'elanger et al. (2004) are summarized in section 3.
In a more recent analysis (section 4) we have used the new version (v. 3.0) of the OSA
and of the calibration files performing the background subtraction with
the latest available background maps to process the whole data set of Table~1.

\section{Preliminary results}
The maps of the Galactic Center shown in Fig.~1
were constructed by summing the reconstructed images of the 571 individual
exposures of the first part of the data set of Table~1,
those taken between 2003 Feb 28 and May 1,
for a total effective exposure time of
about 8.5~$\times$~10$^{5}$~s at the position of \sgra.
In these signal-significance maps of the central two degrees
of the Galaxy
where ten contour levels mark iso-significance
linearly from about $4\sigma$ up to $15\sigma$,
we can see what appear to be six distinct sources:
1E 1740.9--2942.7, KS 1741--293, A 1742--294,
1E 1743.1--2843, SLX 1744--299/300, whose nominal positions
are marked by crosses,
and a source coincident with the radio position of \sgra.
Of these sources, 1E 1740.7--2942 is a black hole candidate and micro-quasar,
KS 1741--293 and A 1742--294 are neutron star Low-Mass X-Ray Binary (LMBX)
burster systems,
SLX 1744-299/300 are in fact two LMXBs separated by only 2.7$'$
and 1E 1743.1--2843 is an X-ray source whose nature is still uncertain (Porquet et al. 2003b).
The 20--40~keV band contours of the central source clearly
peak, with a maximum of $8.7\sigma$, at the \sgra position
but are elongated towards GRS 1741.9--2853.
This suggests some contribution to the emission from this
transient neutron star LMXB burster system
observed to have returned to an active state in 2000 \cite{mun03},
but the elongation could also be due to an uncorrected background structure.
The central source is also marginally visible in the 40--100 keV band
at a level of $4.7\sigma$.
\\
The position and flux of the central excess
in the 20--40~keV map were determined
by fitting the peaks with a function
approximating the instrument's Point Spread Function (Gros et al. 2003)
in two different ways:
(1) all the emission is attributed to one source
and is fitted as such to determine its peak height and position,
(2) the emission is attributed to two sources:
a new source and GRS~1741.9--2853, whose position is then fixed.
Both of these involve a simultaneous fit of all the other sources listed above.
In the first case,
we obtain a source position of
R.A.(J2000.0)=$\mathrm{17^{h}45^{m}22^{s}\hspace{-3pt}.5}$,
decl.(J2000.0)=$-28^{\circ}58'17''$,
and a flux of
about 5.4 mcrab or $(3.21 \pm 0.36) \times$ 10$^{-11}$~ergs~cm$^{-2}$~s$^{-1}$.
In the second,
the position is
$\mathrm{17^{h}45^{m}38^{s}\hspace{-3pt}.5}$,
$-29^{\circ}01'15''$,
and the flux is  about 3.2 mcrab or
$(1.92 \pm 0.36) \times$~10$^{-11}$ergs~cm$^{-2}$~s$^{-1}$.
The central source's 40--100~keV peak
position is in very good agreement with the one determined
using the second method outlined above, and since there is clearly no
visible contribution from a neighboring source,
the 40--100~keV flux was extracted at that position
giving an estimated flux of
$(1.86 \pm 0.40) \times 10^{-11}$~ergs~cm$^{-2}$~$s^{-1}$
($\sim 3.4$~mCrabs).
The fluxes were determined using a standard spectral shape for the
Crab \cite{bar94} and the derived 20--100~keV luminosity
at 8~kpc is $(3.0 \pm 0.4) \times 10^{35}$~erg~s$^{-1}$.
The estimated uncertainty on the position is of about 4$'$
for a detection at the significance level of 8.7$\sigma$
in images still dominated by systematic noise.
These positions are respectively 4.6$'$ and
0.9$'$ from the radio position of \sgra,
within the uncertainties.
\\
The hardness ratio (HR) --- ratio of the count rate in the
high-energy band over that in the low-energy band ---
for the detected excess is $0.90 \pm 0.20$.
As a possible indication of the nature of the detected excess,
we can compare the values of the HR to the two brightest
sources in the field.
The BH candidate 1E~1740.7--2942 has a HR of $1.20 \pm 0.03$,
and the neutron star LMXB KS~1741--293 has a HR of $0.89 \pm 0.08$.
We also found that the source presented some level of
variability, although this result was somehow
hampered by the presence of residual noise
due to uncorrected background structures.
In particular a peak at 5$\sigma$ over the
average was detected in the source light curve on April 6.
This last result has not been confirmed by the most recent
analysis of the data.
\\
As discussed by B\'elanger et al. (2004)
the central excess is not compatible with a simple extrapolation at high energies
of the total X-ray diffuse and point-source flux as observed by X-ray instruments
within 10$'$ from the center.
We concluded that it is due to a hard source,
not identified with the well known high energy sources of the region,
and which was therefore named IGR J1745.6--2901.
\begin{figure}
\centering
\includegraphics[width=0.9\linewidth]{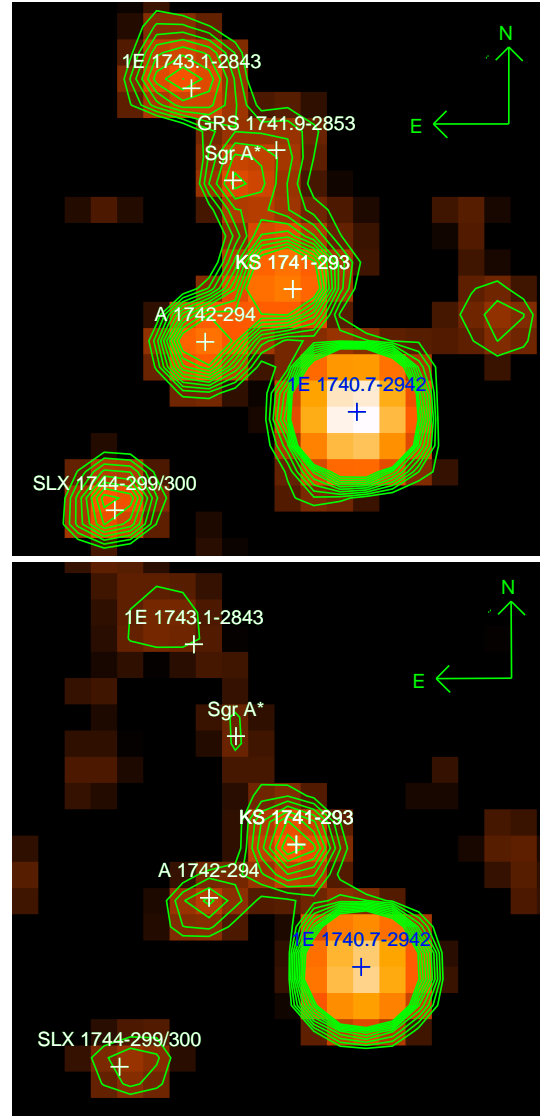}
\caption{
Reconstructed IBIS/ISGRI images of the
central $2^{\circ} \times ~2^{\circ}$ region around the
Galactic Nucleus
        in the energy ranges 20--40~keV (top)
        and 40--100~keV (bottom), using the Feb-March 2003 data
        and preliminary analysis tools.
        Each image pixel size is equivalent to about 5$'$.
        Ten contour levels mark iso-significance
        linearly from about 4$\sigma$ to 15$\sigma$.
Six well known high energy sources are indicated along with
the position of \sgra\ \cite{bel04}
\label{GCINT2040-4010}}
\end{figure}
\section{Recent results}
The whole set of IBIS/ISGRI data collected in the first part of the
2003 were re-analyzed using the new version of the INTEGRAL OSA,
with updated calibration files and lookup tables, and including
the background correction. In addition we performed a preliminary analysis
of the new data collected from August 2003 (see Table~1).
The derived images were much cleaner and the residual noise greatly reduced.
We combined the data in 2 sets of equivalent effective exposure
on the Galactic Nucleus.
An image of the Galactic Center region in the 20-40 keV band was obtained from
all the GCDE and ToO data (Fig.~2).
Using the data of the observation specifically performed to study the
Galactic Center (marked as GO in Table~1) we produced an equivalent image reported in Fig.~3.
Both independent sets of data clearly show the presence of a relevant
excess at the \sgra\ position thus confirming the results reported
by B\'elanger et al. (2004).
IGR J1745.6--2901 was detected at a level of 20$\sigma$ in Fig. 2 and at 27 $\sigma$
in Fig.~3. The combined data provide a signal at the GN of about 35$\sigma$.
Some of the other sources were seen to vary significantly. For example
the other closest source to the center (KS~1741-293) disappeared during the
GO observation (Fig. 3), but IGR J1745.6--2901 was still clearly present.
A fit of the reconstructed images with the PSF of the IBIS/ISGRI telescope
for the six sources detected provided a position of
IGR J1745.6--2901 which is offset from the \sgra\ radio position
by only 52$''$ for the data of Fig.~2 and 48$''$ for the data of Fig.~3.
These offsets are smaller than the expected uncertainty in source location.
\begin{figure}
\centering
\includegraphics[width=0.9\linewidth]{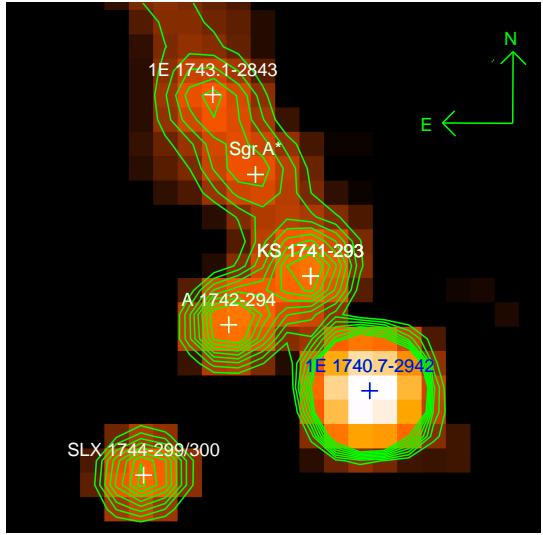}
\caption{
Reconstructed IBIS/ISGRI images of the Galactic Center region
in the 20-40 keV obtained using all data of the 2003 GCDE and ToO
observations and more recent versions of the analysis tools.
Height countours indicate significance levels from 10 to 35$\sigma$.
Five well known high energy sources are indicated along with
the position of Sgr~A$^*$.
\label{Int-gcde03}}
\end{figure}
\begin{figure}
\centering
\includegraphics[width=0.9\linewidth]{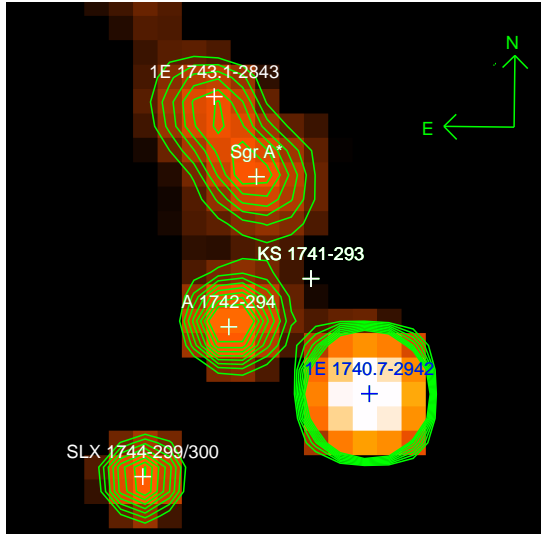}
\caption{
Reconstructed IBIS/ISGRI images of the Galactic Center region
in the 20-40 keV obtained using data of the dedicate observation
of September 2003.
Height countours indicate significance levels from 10 to 35$\sigma$.
Five well known high energy sources are indicated along with
the position of Sgr~A$^*$.
\label{Int-sungold}}
\end{figure}
\\
From the whole data set of Table~1 we obtained the 20-40 keV
light curve reported in Fig. 4. The source flux cumulated in time bins of 1 day
is represented as a function of the universal time.
The source flux appears rather stable over the year with no apparent
large flares.
The average flux was (4.5 $\pm$ 0.12) mCrabs in the 20--40 keV band
and (2.5 $\pm$ 0.23) mCrabs in the 40--60 keV one.
More detailed variability studies will be reported elsewhere.
However we do not confirm the detection of a flare during
the observations of April 6 and the 5$\sigma$ excess
we reported previously \cite{bel04} is probably due to
an uncorrected background feature.
The source GRS 1741.9--2853 also does not appear significant in the
combined images and is not needed to fit the data of Fig.~2 and Fig.~3.
As a consequence, the low energy band flux is more important
than estimated previously, which indicates a softer spectrum
for IGR~J1745.6--2901 than estimated before.
\begin{figure}
\centering
\includegraphics[width=1.0\linewidth]{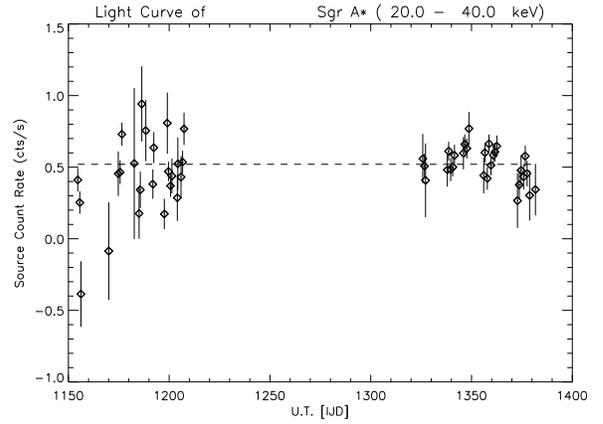}
\caption{IBIS/ISGRI 20-40 keV light curve of IGR J1745.6--2901
from all data collected by INTEGRAL where the source was coded
at $>$~40$\%$. Data points were binned in time intervals of 1 day.
The universal time is in units of IJD~=~MJD-51544 days.
\label{sgrlc}}
\end{figure}
\section{Discussion}
The new analysis of the 2003 INTEGRAL data on the galactic
nucleus confirms the preliminary result of a detection
of a hard source at the position of the Sgr~A complex \cite{bel04}.
The position of this source is offset from the position of \sgra\ by
less then 1$'$ compatible with a 90$\%$ confidence level error radius
of 1.3$'$ for a source at 20$\sigma$ (Gros et al. 2003).
At the moment we cannot associate this excess unambiguously to
\sgra\ or other sources of the region and it is designated as a new INTEGRAL
source, IGR~J1745.6--2901.
As discussed in B\'elanger et al. (2004), IGR~J1745.6--2901 cannot
be explained by the simple extrapolation at high energy
of the average total diffuse and point-source flux observed by the X-ray instruments
within 10$'$ from the center.
On the other hand a few X-ray sources have been
detected in the past within a few arcmin from \sgra\ and they could
contribute to the emission if they were in high/hard state during the INTEGRAL observations.
In particular, the ASCA transient AX J1745.6--2901, compatible in position
with  IGR~J1745.6--2901, was observed to be
bright with Chandra in June 2003 (F. K. Baganoff, private communication) 
and could provide an important contribution
to the excess seen at energies above 20 keV.
\\
Some contribution to the observed high energy emission could also come from
non thermal X-ray filaments observed in the nuclear region
with Chandra and XMM-Newton (Sakano et al. 2003) 
or other non-thermal sources like Sgr~A East.
This option is particularly relevant considering the presence in the region
of an unidentified (GeV) 
gamma-ray EGRET source, 3EG J1746--2851, \cite{may98, har99} and the recent detections
of significant TeV emission from the galactic center with Whipple \cite{kos04}
and Cangaroo-II \cite{tsu04}. All these very high energy sources appear
compatible with a non-variable, point-like source at the GC and could be 
related to IGR J 1745.6--2901.
\\
A detailed analysis of all INTEGRAL data available on the Galactic Nucleus
is in progress and will
provide better constraints on the position, spectral shape, variability properties
and on the possible multiple nature of IGR~J1745.6--2901.
\\
We also expect to further constrain these results by
the simultaneous observations of \sgra\ in gamma-rays, X-rays, and at
infrared and radio wavelengths.
Such observations have been planned, by a large collaboration of
astronomical laboratories, for 2004,
driven by an approved XMM-Newton large project dedicated to the study of
the X-ray flares of \sgra.
The program includes simultaneous observations of the GN with HESS, INTEGRAL,
VLT, HST, VLA and other radio, mm and sub-mm ground based observatories
during part of the 550~ks observing program of XMM-Newton.
Such a program will allow to search for correlated variability
of the \sgra\ emission in different energy domains.
The measure of the broad band spectrum of the flares and its evolution
will allow to constrain the models of the physical processes and
emission mechanisms taking place around the supermassive black hole
at the center of our galaxy.

\section*{Acknowledgements}
Based on observations with INTEGRAL,
an ESA project with instruments and science data centre funded by ESA
member states (especially the PI countries:
Denmark, France, Germany, Italy, Switzerland, Spain), Czech Republic and Poland,
and with the participation of Russia and the USA.
GB and MF acknowledge financial support from
the French Space Agency (CNES). The authors
thank the INTEGRAL project team at ESA for the support
during the phases of the programme.

\end{document}